\documentclass[twocol]{ametsocV5}

\usepackage{amsmath,amsfonts,amssymb,bm}
\usepackage{mathptmx}
\usepackage{newtxtext}
\usepackage{newtxmath}
\usepackage{cancel}
\usepackage{cuted}

\title{A conservation theorem for the $f$~plane}

\authors{Christian E. Buckingham\correspondingauthor{Christian E. Buckingham, christian.buckingham@gmail.com}}

\affiliation{{Universit\'{e} de Bretagne Occidentale, CNRS, IRD, Ifremer, Laboratoire d'Oc\'{e}anographie Physique et Spatiale, {IUEM}, Plouzan\'{e}, France \\
British Antarctic Survey, Cambridge, United Kingdom}}

\abstract{Ertel's potential vorticity theorem is essentially a clever combination of two conservation principles. The result is a conserved scalar $q$ that accurately reflects possible vorticity values that fluid parcels can possess and acts as a tracer for fluid flow. While true at large scales in the ocean and atmosphere, at increasingly smaller scales and in sharply curved fronts, its accuracy breaks down. This is because Earth's rotation imparts angular momentum to fluid parcels and the conservation of absolute angular momentum $L$ restricts the range of centripetal accelerations possible in balanced flow; this correspondingly restricts vorticity. To address this discrepancy, we revisit Ertel's original derivation and obtain a new conserved scalar $Lq$ that more properly reflects the behavior of fluid parcels at these small horizontal scales. Application of the theorem is briefly discussed, with an emphasis on better understanding oceanic submesoscale and polar mesoscale flows.}

\begin{document}

\maketitle

%
%
%


\section{Introduction} \label{SectionIntroduction}

Ocean dynamics at small-scales fronts have garnered considerable attention in recent years. This attention has been evident in both observational and modelling portions of the community. New advancements in \textit{observing} systems--including those from floats \citep{dasaro2011dissipation}, gliders \citep{thompson2016instabilities,plessis2019submeso}, and long-range surface vehicles such as SailDrones \citep{Gentemann2020}--have increased our capability to resolve small-scale phenomena. The result is that velocity and density gradients at horizontal scales between $1$ and $10$~km--previously only inferred from spacecraft \citep[\textit{e.g.}][]{flament1985front,scully1986navy,munk2000spirals} and long-term moored measurements \citep[\textit{e.g.}][]{Bane:1989aa,lilly2002vortex,buckingham2016seasonality}--are now becoming resolved in targeted studies \citep[\textit{e.g.}][]{thomaslee2005,dasaro2011dissipation,thomas2013symmetric,adams2017smiles_alt,garabato2019dynopo}. At the same time, computational resources have increased at an exponential rate, permitting scientists the ability to realistically \textit{simulate} dynamics at these fine scales. At present, numerical models are capable of providing realistic ocean simulations for the globe at horizontal resolutions approaching {$1$~km} ({https://data.nas.nasa.gov/ecco/data.php}). Within nested regional configurations, horizontal grid resolutions of {$100$~m} are possible \citep{onken2020baltic}, with the result that oceanic phenomena with $e$-folding scales of several hundred meters can be resolved.

Oceanic flows at these small spatial scales are commonly referred to as \textit{submesoscale processes} \citep{thomas2008submesoscale,mcwilliams2016review} in order to distinguish them from larger-scale counter-parts, referred to as \textit{mesoscale processes}. At mid-latitudes, these terms correspond to horizontal scales smaller than $10$~km (submesoscale) and larger than $30$~km (mesoscale), where the transition between these scales is roughly defined by the first-mode, baroclinic deformation radius $R_{d}$ \citep{Chelton:2001aa,smith2007bci}. However, at high latitudes, $R_{d}$ approaches $1$-$10$~km \citep{timmermans2008arctic,nurser2014arctic} such that assigning absolute scales to these phenomena is problematic. In the ocean community, this has motivated a dynamical definition for this class of fluid motion.

\subsection{Dynamical definition of the oceanic submesoscale}

Processes within the oceanic submesoscale regime are dynamically characterized by relative vorticity ${\zeta = (\nabla \times \mathbf{u})\cdot \mathbf{\hat{k}}}$ values that rival the vertical component of Earth's vorticity~$f = 2\mathbf{\Omega} \cdot \mathbf{\hat{k}}$ \citep{thomas2008submesoscale,mcwilliams2016review}. Here, $\mathbf{u}$ is velocity, $2\mathbf{\Omega}$ is planetary vorticity, and $\mathbf{\hat{k}}$ is the vertical unit vector. This contrasts with that found for mesoscale processes, where $f$ is typically an order of magnitude greater than relative vorticity, corresponding to quasi-geostrophic (QG) flow.

Additionally, the submesoscale regime is typically characterized by enhanced vertical shears or, equivalently, as a consequence of thermal wind balance (TWB) ${\partial_{z}\mathbf{u}_{h} = \frac{1}{f} \mathbf{\hat{k}} \times \nabla_{h}b}$, enhanced horizontal buoyancy gradients. In Cartesian coordinates oriented relative to the front, we write this balance as $f \partial_{z}\overline{v} = \partial_{x}\overline{b}$, where ${M^2 = \partial_{x}\overline{b}}$ denotes the mean cross-frontal buoyancy gradient. In these expressions, $\mathbf{u_{h}}$ is the horizontal velocity, $\overline{v}$ denotes the mean velocity in the along-front direction, $\partial_{z}\overline{v}$ is the mean vertical shear, $b = -g\rho/\rho_{o}$ denotes buoyancy ($g$ is gravity, $\rho$ is density, and $\rho_{o}$ is a reference density), and $x$ and $y$ are cross-front and along-front coordinates, respectively.

Finally, within boundary layers such as the ocean surface or bottom boundary layers, vertical stratification $N^2 = \partial_{z}\overline{b}$ is often reduced to an extent that fluid parcels within the submesoscale dynamical regime have a greater propensity for vertical motion. These two characteristics are succinctly quantified by the gradient Rossby number (${\mathrm{Ro} = \overline{\zeta}/f}$) and gradient Richardson number (${\mathrm{Ri} = N^2/|\partial_{z}\overline{v}|^2}$), both of which have values approaching $1.0$ within the oceanic submesoscale regime \citep{thomas2008submesoscale,mcwilliams2016review}. In contrast, processes within the quasi-geostrophic regime are characterized by reduced relative vorticity ($\mathrm{Ro} \ll 1$) and elevated stratification and/or reduced vertical shear ($\mathrm{Ri} \gg 1$).

\subsection{Accounting for centripetal accelerations} \label{SectionRefinedDefinition}

It is common to assume that the mean flow within fronts is in approximate geostrophic and hydrostatic balance--\textit{i.e.}~TWB. Additionally, the effect of viscous forces have been considered \citep{mcwilliams2016review}. These are reasonable approximations for density fronts with horizontal scales larger than $R_{d}$ \citep{pedlosky1987gfd}. However, at increasingly smaller scales, the momentum balance can easily shift from geostrophic to cyclogeostrophic balance, reflecting the growing importance of centripetal accelerations. Indeed, vortex generation by baroclinic or barotropic instabilities leads to such a balance. Together with a hydrostatic assumption, this implies a gradient wind balance (GWB): $(f + 2\overline{v}/r) \partial_{z}\overline{v} = \partial_{r}\overline{b}$. 
[For a vector representation of this balance, see \citet{McWilliams1985balance} or \citet{grooms2015submesoscale}.] Factoring out the Coriolis parameter $f$ from the quantity in parentheses immediately leads to a nondimensional parameter which quantifies the impact of centripetal accelerations on the vertical shear: ${\mathrm{Cu} = 2\overline{v}/(fr)}$. This ``curvature'' number also scales with the ratio of centripetal to Coriolis accelerations \citep{shakespeare2016curv}. In the expression above, $r$ is the cross-front coordinate, such that $M^2 = \partial_{r}\overline{b}$ is the radial gradient buoyancy gradient and implicitly contains information regarding frontal curvature. For clarity, we note that ${\mathrm{Cu} > 0}$ for cyclonic curved fronts and ${\mathrm{Cu} < 0}$ negative for anticyclonic curved fronts\footnote{In vortices, $r$ is the distance from the vortex center and is everywhere positive, while $\overline{v} > 0$ for cyclones and $\overline{v} < 0$ for anticyclones. In meandering baroclinic frontal flows, we can replace $r$ with a signed radius of curvature $R$ so long as the along-front flow $\overline{v} > 0$.}. Moreover, in the limit $\mathrm{Cu} \rightarrow 0$ one recovers TWB. GWB is therefore descriptive of curved fronts and vortices, and yet includes TWB as a limiting case. This motivates a slightly refined definition of the oceanic submesoscale, consistent with the former but where curvature effects are explicitly accounted for. In this study, we define the oceanic submesoscale as being a dynamical regime in which the mean flow is in approximate hydrostatic and cyclogeostrophic balance (\textit{i.e.}~GWB), permitting gradient Rossby, Richardson, and curvature numbers of order-one: $(\mathrm{Ro},\mathrm{Ri},\mathrm{Cu}) \sim 1$.

\subsection{Motivation}

In a previous study \citep{buckingham2020part1,buckingham2020part2}, it was suggested that a unique conservation principle may be present within highly curved fronts and vortices (\textit{i.e.}~``vortex flow'') on the $f$-plane. Moreover, this principle was invoked when proposing a mechanism for the evolution of small-scale (\textit{i.e.}~submesoscale and polar mesoscale) vortices in the ocean. The implication was that fluid parcels within curved baroclinic fronts and vortices do not simply conserve the Ertel potential vorticity (PV) \citep{ertel1942}, and therefore undergo vortex stretching and tilting to conserve this quantity. Rather, fluid parcels adjust barotropic and baroclinic components of another scalar quantity,\footnote{\citet{buckingham2020part1,buckingham2020part2} suggested that the generalized Rayleigh discriminant $\Phi = 2Lq/r^2$ \citep{kloosterziel2007inertial} was conserved following fluid parcels in highly curved fronts and vortices. However, as demonstrated below, this statement is incorrect: it is $Lq$ or ${r^2\Phi}$ that is conserved following fluid parcels. This difference is critical because it implies cross-frontal motion will modify the stability ``seen'' by fluid parcels.} which is proportional to the product of the Ertel PV ($q$) and the vertical component of absolute angular momentum ($L$). If true, the fact that this additional term $L$ enters the conserved scalar provides an added constraint to the problem, thereby making parcel motion within highly curved baroclinic flows differ from those in which PV alone is the conserved scalar. As will be demonstrated below, this places constraints on vorticity.

The purpose of this manuscript is therefore to provide a rational argument for the statement that {``the product of the absolute angular momentum and Ertel PV is conserved following fluid parcels.''} Moreover, we wish to assess under which conditions such a statement is true. In doing so, we lay a more formal foundation for the analysis of submesoscale baroclinic flows in which centripetal accelerations cannot be neglected. 

\subsection{Outline}

The outline of this study is as follows. We first derive a conservation theorem for the new scalar quantity ${Lq}$ (section~\ref{SectionTheorem}). This derivation closely follows that of \citet{ertel1942} but it includes a brief presentation of absolute angular momentum conservation--a topic typically neglected in oceanographic studies. Second, the application of the resulting theorem is briefly discussed in section~\ref{SectionDiscussion} and its limitations mentioned in section~\ref{SectionLimitations}. We conclude the study in section~\ref{SectionConclusions}.

\section{Derivation} \label{SectionTheorem}

\begin{figure}[t]
\begin{centering}
\noindent\includegraphics[width=18pc,angle=0]{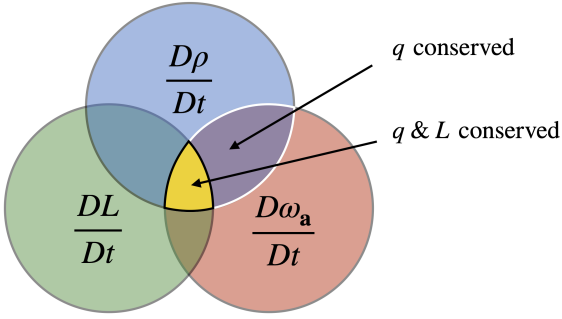}
  \caption{Venn diagram conceptually depicting the intersection of three conservation principles: absolute vorticity $\mathbf{\omega_{a}}$, density $\rho$, and the vertical component of absolute angular momentum ${L}$. Here, the sources and sinks in each equation are implied. While \citet{ertel1942} focused on the intersection of density and vorticity conservation (\textit{i.e.}~$q$ conserved), this study examines the intersection of density, vorticity, and absolute angular momentum conservation (\textit{i.e.}~$q$ \& $L$ conserved). We emphasize that conservation equations for $\rho$, $\mathbf{\omega_{a}}$, and $L$ are not independent, thereby making this intersection possible.}\label{f1}
\end{centering}
\end{figure}

Ertel's~(1942) PV theorem is essentially an intersection of two conservation principles: vorticity and density. It is logical to presume that the inclusion of a third conservation principle together with its corresponding conditions could permit a new vorticity theorem subject to these additional limitations. This is the central concept behind the present study, where the third conservation law is provided by the evolution equation for \textit{absolute angular momentum} (Figure~\ref{f1}).

\subsection{Governing equations} \label{SectionGoverningEqn}

The equations of motion describing the balance of forces per unit mass of a fluid parcel within a rotating reference frame are \citep{batchelor1967intro,pedlosky1987gfd,cushmanroisin1994intro}
\begin{equation} \label{EquationMomRotatingSimple}
\frac{D\mathbf{u}}{Dt} + 2\mathbf{\Omega} \times \mathbf{u} = 
 -\frac{1}{\rho}\nabla p
 + \underbrace{\mathbf{g^{*}} + \mathbf{a_{c}}}_{\mathrm{}~\mathbf{g}}
 + \frac{\mathscr{F}}{\rho},
\end{equation}
where it is understood that all terms are evaluated within the rotating reference frame. Here, ${D/Dt = \partial_{t} + \mathbf{u} \cdot \nabla \mathbf{u}}$ denotes the material or substantial derivative, $\mathbf{r}$ is the position vector, $\mathbf{\Omega}$ is the angular rotation rate ({$|\mathbf{\Omega}|~=~2\pi/$day~$\approx~7.22~\times 10^{-5}$~s$^{-1}$} for Earth) and assumed to be constant, $2\mathbf{\Omega} \times \mathbf{u}$ is the Coriolis acceleration, ${\mathbf{a_{c}} = -\mathbf{\Omega} \times (\mathbf{\Omega} \times \mathbf{r}) = |\mathbf{\Omega}|^2 \mathbf{r_{\perp}}}$ is the centrifugal acceleration due to the rotation of the reference frame, $\rho$ is density, $p$ is pressure, $\mathbf{g^{*}}$ is the acceleration due to gravity, and $\mathscr{F}$ denotes the frictional force.

\begin{figure}
  \centerline{\includegraphics[width=22pc,angle=0]{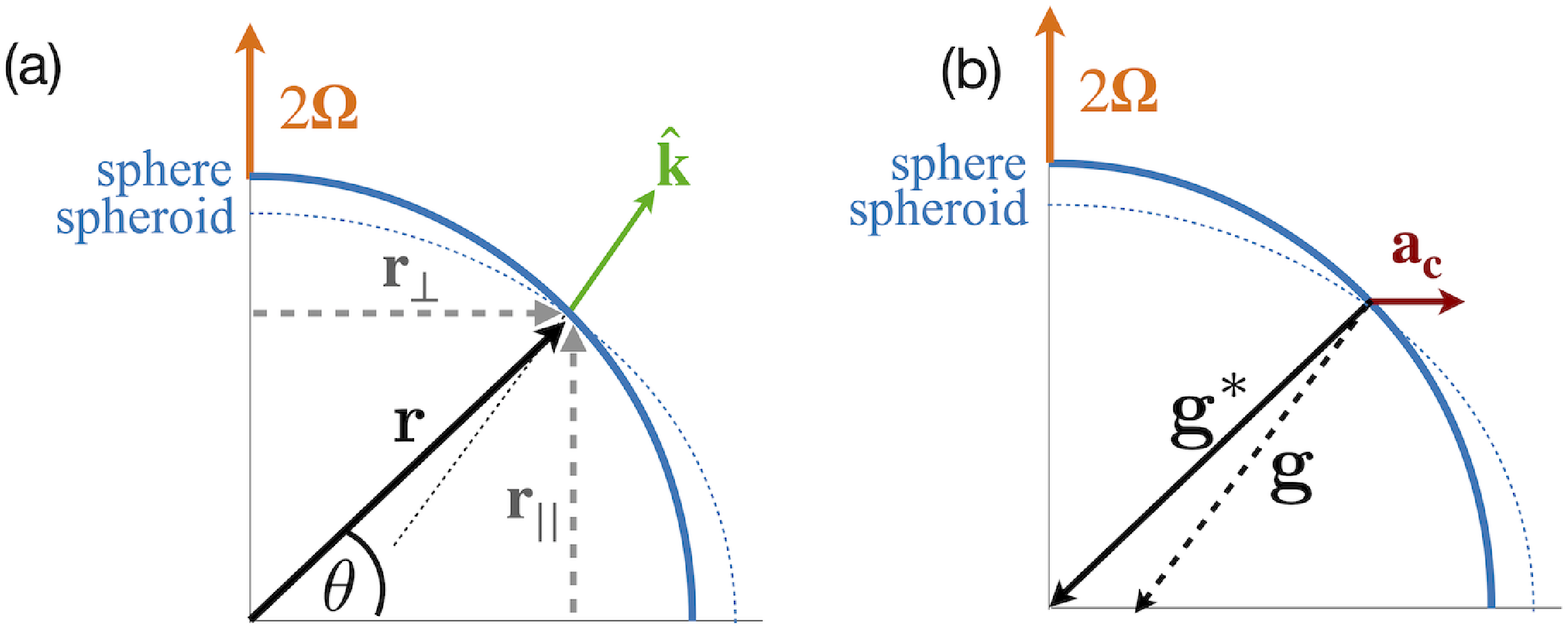}}
  \caption{Illustration of vectors present within the equations of motion on the sphere (cf.~Equation~\ref{EquationMomRotatingSimple}) and $f$~plane approximation (cf.~Equation~\ref{EquationMomRotatingFplaneVec}). In (a), we depict planetary vorticity $2\mathbf{\Omega}$ (orange), the position vector $\mathbf{r}$ (heavy black), components of the position vector $\mathbf{r_{\perp}}$ and $\mathbf{r_{||}}$ (gray), and vertical unit vector $\mathbf{\hat{k}}$ (green). In (b), we depict the gravitational vector $\mathbf{g^{*}}$ (black), the centrifugal acceleration vector ${\mathbf{a_{c}} = -\mathbf{\Omega} \times (\mathbf{\Omega} \times \mathbf{r}) = |\mathbf{\Omega}|^2\mathbf{r_{\perp}}}$ (red), and the vector resultant, or effective gravity ${\mathbf{g} = \mathbf{g^{*}} + \mathbf{a_{c}}}$ (dashed black). We also illustrate the surface of Earth as represented by a sphere (solid blue) and oblate sphere, or spheroid (dashed blue). The unit vector $\mathbf{\hat{k}}$ is anti-parallel to $\mathbf{g}$ and, therefore, approximately perpendicular to the spheroid's surface.}
\label{f2}
\end{figure}

It is customary to combine centrifugal and gravitational accelerations into a resultant acceleration~${\mathbf{g} = \mathbf{g^{*}} + \mathbf{a_{c}}}$, or \textit{effective gravity}. The resultant is then approximately perpendicular to geopotential surfaces and, hence, oriented vertically\footnote{Local changes to the gravitational potential, for example, due to irregular topography or seamounts, will perturb $\mathbf{g^{*}}$ from its mean direction.} \citep{cushmanroisin1994intro}. For clarity, we illustrate planetary vorticity, gravitational acceleration, gravity, and centrifugal acceleration vectors (Figure~\ref{f2}). Mass conservation is given by the continuity equation
\begin{equation} \label{EquationContinuityFull}
\frac{\partial \rho }{\partial t } + \mathbf{\nabla} \cdot (\rho \mathbf{u}) = 0.
\end{equation}
An equation of state is necessary to relate $\rho$ to known or measured variables. In the ocean, this is a complex function of temperature, salinity, and pressure. For simplicity, we will assume the density is known perfectly.

\begin{figure}[t]
\begin{centering}
\noindent\includegraphics[width=20pc,angle=0]{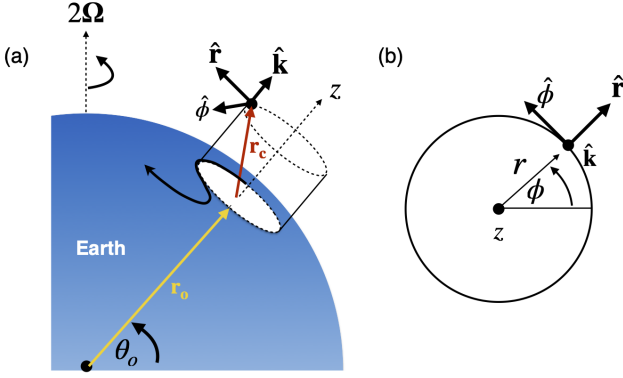}
  \caption{A cylindrical coordinate system on an $f$-plane at latitude $\theta = \theta_{o}$: (a) perspective view and (b) plan view, illustrating the orthogonal unit basis $(\mathbf{\hat{r}},\mathbf{\hat{\phi}},\mathbf{\hat{k}})$, position vector ${\mathbf{r_{c}} = (r,\phi,z)}$ (red), where the angle $\phi$ is defined relative to an eastward direction, and a vector $\mathbf{r_{o}}$ (yellow) which helps define the origin of the cylindrical coordinate system. Although not shown, the velocity is ${\mathbf{u}=(u,v,w)}$ and its components point in $\mathbf{\hat{r}}$, $\mathbf{\hat{\phi}}$, and $\mathbf{\hat{k}}$ directions, respectively.} \label{f3}
\end{centering}
\end{figure}

\subsubsection*{Restriction to small horizontal scales} \label{SectionCoordinateSys}

The corresponding equations of motion valid under an approximation of a constant rotation rate of the reference frame (\textit{i.e.}~$f$-plane) are \textit{formally obtained} by expressing Equation~\ref{EquationMomRotatingSimple} in spherical coordinates, scaling the equations of motion, and discarding terms multiplied by $|ds|/R_{e} \ll 1$ or smaller, where $|ds| = R_{e}d\theta$ denotes a meridional arc length and $R_{e}$ is the mean radius of Earth \citep{Grimshaw1975beta}. The result is a vectorized set of equations comparable to Equation~\ref{EquationMomRotatingSimple} except where $2\mathbf{\Omega} \times \mathbf{u}$ is now evaluated at a specific latitude $\theta_{o}$:
\begin{equation} \label{EquationMomRotatingFplaneVec}
\frac{D\mathbf{u}}{Dt} + 2\mathbf{\Omega_{o}} \times \mathbf{u} = 
 -\frac{1}{\rho}\nabla p 
 + \mathbf{g}
 + \frac{\mathscr{F}}{\rho}.
\end{equation}
Note that making the $f$~plane approximation does not alter the continuity equation (cf.~Equation~\ref{EquationContinuityFull}).

For later reference, we detail the above terms in cylindrical polar coordinates. In cylindrical coordinates, where the triad of orthogonal unit vectors $(\mathbf{\hat{r}},\mathbf{\hat{\phi}},\mathbf{\hat{k}})$ point in radial, azimuthal, and vertical (upward) directions, respectively, we denote the position vector by ${\mathbf{r_{c}} = (r,\phi,z)}$ and velocity by ${\mathbf{u} = (u,v,w)}$ (Figure~\ref{f3}). The material derivative is then ${D/Dt = \partial_{t} + \mathbf{u} \cdot \nabla = \partial_{t} + u\partial_{r} + (v/r)\partial_{\phi} + w\partial_{z}}$. Finally, the frictional force is $\mathscr{F} = (F_{r},F_{\phi},F_{z})$ and effective gravity is $\mathbf{g} = (0,0,-g)$. Note that the choice of a cylindrical coordinate system on the $f$~plane complicates expression of $\mathbf{\Omega_{o}}$ owing to its variation with azimuth angle $\phi$. \textbf{To retain generality in our derivation below, we use the vector form of Equation~\ref{EquationMomRotatingFplaneVec} together with the full Coriolis vector $\mathbf{\Omega_{o}}$}. 

\subsection{Absolute vorticity conservation}

The following two sections can be found elsewhere \citep[\textit{e.g.}][]{pedlosky1987gfd,muller1995,vallis2017text}, but are repeated here for completeness. Equation~\ref{EquationMomRotatingFplaneVec} can be recast in terms of the absolute vorticity or ``Stokes'' form \citep{batchelor1967intro}:
\begin{equation} \label{EquationMomRotatingVorticity1}
\frac{\partial{\mathbf{u}}}{\partial{t}} + \mathbf{\omega_{a}} \times \mathbf{u}  =  -\frac{1}{\rho}\nabla p 
 + \nabla \left [ \mathbf{g} \cdot \mathbf{r_{c}} - (\mathbf{u}\cdot\mathbf{u})/2 \right ] 
 + \frac{\mathscr{F}}{\rho},
\end{equation}
where ${\omega_{a} = \nabla \times \mathbf{u_{a}} = 2\mathbf{\Omega_{o}} + \nabla \times \mathbf{u} = 2\mathbf{\Omega_{o}} + \mathbf{\omega}}$ is absolute vorticity (\textit{i.e.}~the sum of relative and planetary vorticity) and ${\mathbf{u_{a}} = \mathbf{u} + \mathbf{\Omega} \times \mathbf{r_{c}}}$ is absolute velocity.
Taking the curl of Equation~\ref{EquationMomRotatingVorticity1} gives
\begin{equation} \label{EquationMomRotatingVorticity2}
\frac{\partial{\mathbf{\omega}}}{\partial{t}} + \nabla \times \left (\mathbf{\omega_{a}} \times \mathbf{u} \right )  =  \frac{\nabla \rho \times \nabla p}{\rho^2} 
 + \nabla \times \left (\frac{\mathscr{F}}{\rho} \right ).
\end{equation}
Using the identity \citep[\textit{e.g.}][]{Riley:2006aa}
\begin{equation}
{\nabla \times \left (\mathbf{A} \times \mathbf{B} \right )} = 
\mathbf{A} \nabla \cdot \mathbf{B}
+ (\mathbf{B} \cdot \nabla )\mathbf{A}
- \mathbf{B} \nabla \cdot \mathbf{A}
- (\mathbf{A} \cdot \nabla )\mathbf{B},
\end{equation}
and noting that
the planetary vorticity is constant,\footnote{This is true regardless of the chosen coordinate system since the vector $\mathbf{2\Omega}$ remains unchanged.} 
one can re-write the vorticity equation as
\begin{equation} \label{EquationVorticityConservation}
\frac{D\mathbf{\omega_{a}}}{Dt} = \mathbf{\omega_{a}}\cdot{\nabla \mathbf{u}} - \omega_{a} \nabla \cdot \mathbf{u} + \frac{\nabla \rho \times \nabla p}{\rho^2} + \nabla \times \left ( \frac{\mathscr{F}}{\rho} \right ).
\end{equation}
This equation states that the absolute vorticity of a fluid element is modified by (i)~shearing motion that tilts or re-orients the vorticity vector or by vortex stretching, (ii)~compressibility, (iii)~baroclinicity, which alters the center of mass relative to that found when density contours and pressure contours are parallel, and (iv)~frictional forces. This is the basis for vortex stretching and tilting interpretations of PV.

Another useful form of the vorticity equation is obtained as follows. One can replace the divergence term in the vorticity equation using Equation~\ref{EquationContinuityFull}, and rewrite the material derivative of absolute vorticity per unit mass as
\begin{equation}
\frac{D}{Dt}\left (\frac{\mathbf{\omega_{a}}}{\rho}\right ) = \frac{1}{\rho} \left [ \frac{D\mathbf{\omega_{a}}}{Dt} - \frac{\mathbf{\omega_{a}}}{\rho}\frac{D\rho}{Dt} \right ],
\end{equation}
allowing one to express Equation~\ref{EquationVorticityConservation} as the conservation of absolute vorticity per unit mass:
\begin{equation} \label{EquationVorticityConservationPerMass}
\frac{D}{Dt}\left (\frac{\mathbf{\omega_{a}}}{\rho}\right ) = \left (\frac{\mathbf{\omega_{a}}}{\rho}\cdot \nabla \right ) \mathbf{u} + \frac{\nabla \rho \times \nabla p}{\rho^3} + \left ( \nabla \times \frac{\mathscr{F}}{\rho} \right )\frac{1}{\rho}.
\end{equation}

\subsection{Conservation of density}

The next step displays the creativity of Ertel.
Following \citet{pedlosky1987gfd} (Ertel assumes $D\rho/Dt = 0$), we write the conservation of a scalar quantity $\lambda$ as 
\begin{equation} \label{EquationDensityConservation}
D\lambda/Dt = \frac{\partial{\lambda}}{\partial{t}} + \mathbf{u}\cdot \nabla \lambda = \Psi.
\end{equation}
Taking the inner product of $\nabla \lambda$ and Equation~\ref{EquationVorticityConservationPerMass}, one obtains
\begin{equation} \label{EquationErtelPvConservation1a}
\nabla \lambda \cdot \frac{D}{Dt}\left (\frac{\mathbf{\omega_{a}}}{\rho}\right ) = 
\nabla \lambda \cdot \left [ \left (\frac{\mathbf{\omega_{a}}}{\rho}\cdot \nabla \right ) \mathbf{u} \right ] 
+ \nabla \lambda \cdot \frac{\nabla \rho \times \nabla p}{\rho^3}
+ \frac{\nabla \lambda}{\rho} \cdot \left ( \nabla \times \frac{\mathscr{F}}{\rho} \right ).
\end{equation}
Incorporating $\nabla \lambda$ into the material derivative on the left-hand-side (LHS),\footnote{This follows from ${\mathbf{A} \cdot \frac{D ( \nabla \lambda)}{Dt} = \mathbf{A} \cdot \nabla \frac{D\lambda}{Dt} - \nabla \lambda \cdot \left ( \mathbf{A} \cdot  \nabla \mathbf{u} \right )}$, where we have used ${\mathbf{A} = {\mathbf{\omega_{a}}}/{\rho}}$.}
we obtain
\begin{equation} \label{EquationErtelPvConservationMain}
\frac{Dq}{Dt} = \frac{D}{Dt}\left ( \frac{\mathbf{\omega_{a}}}{\rho} \cdot \nabla \lambda \right ) = \frac{\mathbf{\omega_{a}}}{\rho}\cdot \nabla \Psi + \nabla \lambda \cdot \frac{\nabla \rho \times \nabla p}{\rho^3} + \frac{\nabla \lambda}{\rho} \cdot \left ( \nabla \times \frac{\mathscr{F}}{\rho} \right ).
\end{equation}
This states that the quantity, ${q = (\mathbf{\omega_{a}}}/\rho)\cdot \nabla \lambda$, is conserved following fluid parcels if the RHS is zero. Choosing, for example, density as our scalar quantity, $\lambda = \rho$, while requiring frictional and diabatic processes to be zero so that the flow is inviscid and density is conserved, we see that all three terms on the RHS~vanish. This is Ertel's (1942) vorticity theorem. More generally, $\lambda$ can be any variable so long as it is a function of density and pressure--\textit{i.e.}~a ``thermodynamic function'' \citep{pedlosky1987gfd}.

\subsection{Absolute angular momentum conservation} \label{SectionAbsAngMom}

One of the contributions of \citet{rayleigh1917} was to demonstrate that, if a vortex is axisymmetric, then the azimuthal momentum equation can be multiplied by $r$ and re-expressed as a conservation equation for the angular momentum per unit mass: ${D/Dt(r\overline{v}) = 0}$, where $\overline{v}$ denotes the azimuthal velocity. Application of this approach to a fluid parcel in a rotating reference frame with constant rotation rate also permits such a rearrangement: ${DL/Dt = 0}$, where $L = r \overline{v} + fr^2/2$ is now the \emph{absolute angular momentum}, and is the sum of relative angular momentum ($r \overline{v}$) and planetary angular momentum imparted by the rotating reference frame. Importantly, the absolute angular momentum of a fluid parcel in a vortex on the $f$~plane \textit{is exactly the same as if the vortex were located at the center of the rotating reference frame}, where $r$ is the magnitude of the position vector \citep{Kloosterziel:1991aa}. \textbf{This motivates the following vector representation.}

We orient our coordinate system so that its origin is at the center of a curved front or vortex (cf.~Figure~\ref{f3}). Taking the cross product of the position vector $\mathbf{r_{c}}$ and each of the terms in Equation~\ref{EquationMomRotatingFplaneVec}, one obtains after some effort
\begin{equation} \label{EquationConservAngMom1}
\frac{D\mathbf{m_{a}}}{Dt} = - \mathbf{\Omega_{o}} \times \mathbf{m}
 - \frac{\mathbf{r_{c}} \times \nabla p}{\rho}
+ \mathbf{r_{c}} \times \mathbf{g}
+ \frac{\mathbf{r_{c}} \times \mathscr{F}}{\rho},
\end{equation}
where ${\mathbf{m_{a}} = \mathbf{r_{c}} \times \mathbf{u_{a}}}$ and $\mathbf{m} = \mathbf{r_{c}} \times \mathbf{u}$ are absolute and relative angular momentum, respectively. Using the definition of absolute velocity, ${\mathbf{u_{a}} = \mathbf{u} + \mathbf{\Omega_{o}} \times \mathbf{r_{c}}}$, we observe that ${\mathbf{m_{a}} =  \mathbf{m} + \mathbf{m_{\Omega}}}$ is the sum of relative angular momentum ${\mathbf{m} = \mathbf{r_{c}} \times \mathbf{u}}$ and planetary angular momentum ${\mathbf{m_{\Omega}} = \mathbf{r_{c}} \times (\mathbf{\Omega_{o}} \times \mathbf{r_{c}})}$ in a manner analogous to absolute vorticity~$\mathbf{\omega_{a}}$.

For our purposes, we wish to isolate the vertical component of absolute angular momentum. We take the inner product of Equation~\ref{EquationConservAngMom1} and the vertical unit vector $\mathbf{\hat{k}}$ to obtain
\begin{equation} \label{EquationConservAngMomVert}
\frac{DL}{Dt} = 
- \left ( \mathbf{\Omega_{o}} \times \mathbf{m} \right ) \cdot \mathbf{\hat{k}}
- \frac{\mathbf{r_{c}} \times \nabla p}{\rho} \cdot \mathbf{\hat{k}}
+ ( \mathbf{r_{c}} \times \mathbf{g} ) \cdot \mathbf{\hat{k}}
+ \frac{\mathbf{r_{c}} \times \mathscr{F}}{\rho} \cdot \mathbf{\hat{k}},
\end{equation}
where we have introduced the notation ${L = \mathbf{m_{a}} \cdot \mathbf{\hat{k}}}$ to denote the vertical component of absolute angular momentum, consistent with the literature \citep{holton1992text,shakespeare2016curv}. Thus, the vertical component of absolute angular momentum $L$ of a fluid parcel is modified by torques due to pressure, gravitation, and friction, as well as a torque produced by Earth's rotation acting on the relative angular momentum $\mathbf{m}$. For cases when $\mathbf{m}$ is not vertical, the latter reduces $L$, tilting the absolute angular momentum vector away from the vertical.

\subsubsection*{Angular momentum conservation on the spheroid}

While our interest is in small-scale fronts and vortices, it is nonetheless helpful to compare the conservation equation above (cf.~Equation~\ref{EquationConservAngMom1}) with that obtained for the oblate sphere \citep[\textit{e.g}][]{barnes1983momentum,piexoto_oort1992physics,bell1994momentum}. This was considered, for example, by \citet{egger2001ang}. In this case, the position vector $\mathbf{r}$ extends from Earth's center to the fluid parcel (Figure~\ref{f2}). Computing the cross product of $\mathbf{r}$ and the more general equations of motion (cf.~Equation~\ref{EquationMomRotatingSimple}), one obtains
\begin{equation} \label{EquationConservAngMom2}
\frac{D\mathbf{m_{a}}}{Dt} =
- \mathbf{\Omega} \times \mathbf{m_{a}}
- \frac{\mathbf{r} \times \nabla p}{\rho}
+ \mathbf{r} \times \underbrace{( \mathbf{g} - \mathbf{a_{c}})}_{\mathbf{g^{*}}}
+ \frac{\mathbf{r} \times \mathscr{F}}{\rho},
\end{equation}
where now the absolute, relative, and planetary angular momentum are given, respectively, by ${\mathbf{m_{a}} = \mathbf{m} + \mathbf{m_{\Omega}}}$, ${\mathbf{m} = \mathbf{r} \times \mathbf{u}}$, and ${\mathbf{m_{\Omega}} = \mathbf{r} \times (\mathbf{\Omega} \times \mathbf{r})}$. Expanding the first term on the RHS and examining only the planetary portion of this term, we see that Earth's rotation induces a torque with magnitude ${|-\mathbf{\Omega} \times \mathbf{m_{\Omega}}| = |\mathbf{r_{||}}| |\mathbf{\Omega}| | \mathbf{\Omega} \times \mathbf{r_{\perp}} | = |\mathbf{\Omega}|^2 |\mathbf{r_{\perp}}||\mathbf{r_{||}}|}$ that is directed eastward.\footnote{Note: ${\mathbf{\Omega} \times \mathbf{r} = \mathbf{\Omega} \times \mathbf{r_{\perp}}}$ and ${\mathbf{A} \times (\mathbf{B} \times \mathbf{C}) = (\mathbf{A} \cdot \mathbf{C} ) \mathbf{B} - (\mathbf{A} \cdot \mathbf{B}) \mathbf{C}}$ allow us to simplify the planetary angular momentum: ${\mathbf{m_{\Omega}} = |\mathbf{r_{\perp}}|^2 \mathbf{\Omega} - |\mathbf{r_{||}}| |\mathbf{\Omega}| \mathbf{r_{\perp}}}$.} Similarly, the torque induced by the centrifugal acceleration has magnitude ${|-\mathbf{r} \times \mathbf{a_{c}}| = |\mathbf{\Omega}|^2 |\mathbf{r_{\perp}}||\mathbf{r_{||}}|}$ but is directed westward. That is, the two terms cancel and Equation~\ref{EquationConservAngMom2} can be written simply as
\begin{equation} \label{EquationConservAngMom2alt}
\frac{D\mathbf{m_{a}}}{Dt} =
- \mathbf{\Omega} \times \mathbf{m}
- \frac{\mathbf{r} \times \nabla p}{\rho}
+ \mathbf{r} \times \mathbf{g}
+ \frac{\mathbf{r} \times \mathscr{F}}{\rho}.
\end{equation}
We now note that Equation~\ref{EquationConservAngMom2alt} is identical to Equation~\ref{EquationConservAngMom1} except where $\mathbf{r_{c}}$ is replaced by $\mathbf{r}$. Thus, while a formal proof remains, we argue that absolute angular momentum is conserved on the $f$~plane similarly as how it is conserved on the sphere.\footnote{\citet{egger2001ang} did not demonstrate this vector cancellation and led him to conclude that angular momentum conservation was different in the $f$~plane approximation than on the spheroid (in the limit of small $|ds|/R_{e}$). We disagree with this statement for the reasons stated above.}  This may be why, for sufficiently small horizontal scales and for balanced (\textit{i.e.}~hydrostatic) flows in which the meridional component of Coriolis is neglected, the volume-integrated, vertical component of absolute angular momentum is approximately conserved \citep[][Fig.~2e]{egger2001ang}.

\subsection{A vorticity theorem for the $f$~plane} \label{SectionCombine}

We are now in a position to combine conservation laws (cf.~Equations~\ref{EquationErtelPvConservationMain} and \ref{EquationConservAngMomVert}). It is simple to show that if $\frac{DA}{Dt} = 0$ and if $\frac{DB}{Dt} = 0$, then $\frac{D}{Dt}(AB) = 0$. Although elementary, this is the logic behind the following step. We multiply Equation~\ref{EquationErtelPvConservationMain} by ${L = \mathbf{m_{a}} \cdot \mathbf{\hat{k}} = (\mathbf{m} + \mathbf{m_{\Omega}} ) \cdot \mathbf{\hat{k}}}$ and add this to $q = ({\mathbf{\omega_{a}}}/{\rho}) \cdot \nabla \lambda$ multiplied by Equation~\ref{EquationConservAngMomVert}. This gives
\begin{eqnarray} \label{EquationConservLq}
\frac{D}{Dt}(Lq) & = & L \left [ \frac{\mathbf{\omega_{a}}}{\rho}\cdot \nabla \Psi + \nabla \lambda \cdot \frac{\nabla \rho \times \nabla p}{\rho^3} + \frac{\nabla \lambda}{\rho} \cdot \left ( \nabla \times \frac{\mathscr{F}}{\rho} \right ) \right ] \nonumber \\
 & + &  q \big [
 - {\left ( \mathbf{\Omega_{o}} \times \mathbf{m} \right )} \cdot \mathbf{\hat{k}}
- \frac{\mathbf{r_{c}} \times \nabla p}{\rho} \cdot \mathbf{\hat{k}} \nonumber \\
& + & \left ( \mathbf{r_{c}} \times \mathbf{g} \right ) \cdot \mathbf{\hat{k}}
+ \frac{\mathbf{r_{c}} \times \mathscr{F}}{\rho} \cdot \mathbf{\hat{k}} \big ].
\end{eqnarray}
where it is emphasized that $\mathbf{r_{c}}$ is the {position vector in the cylindrical coordinate system} and $\mathbf{g}$ is directed anti-parallel to the unit vector $\mathbf{\hat{k}}$ (Figure~\ref{f3}). 
This equation states that the scalar quantity $Lq$ is conserved following fluid parcels on the $f$~plane if, for non-zero $L$ and $q$, all of the following conditions are met:
\begin{enumerate}
\item{density is conserved ($\Psi = 0$)}
\item{the fluid is inviscid ($\mathscr{F} = 0$)}
\item{
\begin{enumerate}
\item{the fluid is barotropic ($\nabla \rho \times \nabla p = 0$) or}
\item{the fluid is baroclinic ($\nabla \rho \times \nabla p \ne 0$) and $\lambda$ is a ``thermodynamic function''}
\end{enumerate}
}
\item{relative angular momentum $\mathbf{m}$ is directed vertically so that ${(\mathbf{\Omega_{o}} \times \mathbf{m}) \cdot \mathbf{\hat{k}} = 0}$}
\item{pressure torques are zero or orthogonal to the vertical so that ${(\mathbf{r_{c}} \times \nabla p ) \cdot \mathbf{\hat{k}} = 0}$}
\item{perturbations in Earth's gravitational field are zero so that ${(\mathbf{r_{c}} \times \mathbf{g} ) \cdot \mathbf{\hat{k}} = 0}$, and}
\end{enumerate}

While the above conservation theorem may find reduced application when compared to Ertel's PV theorem owing to the number of aforementioned conditions, several simplifications are possible. For adiabatic and inviscid baroclinic flows, selecting $\lambda$ as a thermodynamic variable (\textit{e.g.}~$\lambda = -\rho g$) satisfies conditions~(1)-(3). For geophysical flows of the type considered here, the flow is nearly two-dimensional such that $\mathbf{m}$ points approximately vertically. In this case, $\mathbf{\Omega_{o}} \times \mathbf{m}$ will be tangent to Earth's surface and condition~(4) can be met. In the absence of geopotential perturbations, condition~(6) is met. Finally, assuming symmetry in the direction of the flow, pressure gradient torques are zero. (This may not be true if perturbations due to baroclinic instability break such symmetry.) However, undulating topography introduces pressure torques. This includes ocean bottom topography or rough ice (\textit{i.e.}~for vortex flows found beneath sea ice or ice shelves). Thus, condition~(5) is met for \textit{azimuthally symmetric flow away from boundaries}. Conditions~(1)-(3) correspond exactly to Ertel's PV theorem, while conditions (4)-(6) ensure conservation of $L$. In conclusion, we have a conservation theorem valid on the $f$~plane that is different than Ertel's PV theorem and yet, at least in curved balanced flows away from boundaries, has the potential to satisfy all of the aforementioned conditions. \textit{If these conditions are met, the product of the vertical component of absolute angular momentum and Ertel PV ($Lq$) is conserved following fluid parcels.}\footnote{It is not clear how best to refer to the quantity $Lq$. We were at first tempted to refer to $Lq$ as a generalized form of PV since fluid parcels have possible vorticity values that are set by the sign of $Lq$ through the stability discriminant $\Phi = 2Lq/r^2$ \citep{buckingham2020part1,buckingham2020part2}. However, the theorem's validity is confined to small horizontal scales such that $Lq$ is not universally conserved. For this reason, the term \textit{submesoscale potential vorticity} might be a suitable alternative. Note: the scalar $\Phi/f = 2Lq/(fr^2) = (1 + \mathrm{Cu})q$ is perhaps a better variable to be named {submesoscale potential vorticity} since it shares the same units as $q$ and applies to straight and curved fronts. One recovers the Ertel PV in the limit $\mathrm{Cu} \rightarrow 0$. In any case, to avoid conflict with the Ertel PV and given its established relationship to angular momentum \citep{rayleigh1917,solberg1936,fjortoft1950vortex}, we adopt the term \textit{potential momentum} below in reference to $Lq$, reflecting that changes in angular momentum can occur as a result of alterations in the baroclinic nature of the fluid.}

\section{Discussion} \label{SectionDiscussion}

It is helpful to consider how such a conservation principle might find practical application. Given the restriction to the $f$~plane, we propose that the conservation theorem will find greatest application in understanding vortex flows away from the tropics in the oceanic submesoscale regime (cf.~section~\ref{SectionIntroduction}\ref{SectionRefinedDefinition}). These will include (1)~polar mesoscale eddies found under ice in the Arctic \citep{DAsaro:1988aa,timmermans2008arctic,zhao2014arctic}, (2)~simulated vortices in the laboratory \citep{stegner2004part1,kloosterziel2007inertial,lazar2013part2}, (3)~hydrothermal vents and convective plumes \citep{helfrich1991plume,dasaro1994plume,legg2001labrador,deremble2016plume}, and (4)~meddies and other mid-latitude coherent vortices formed through interactions of currents with topography, cyclogeostrophic adjustment of buoyant waters, and vertical convection through buoyancy fluxes in boundary layers \citep{mcdowell1978meddy,mcwilliams1985scv,riser1986scv,Bane:1989aa,kostianoy1989scv,lilly2002vortex,bosse2016scv,Meunier:2018aa}. We note that it might also find relevance in understanding (5) parcel motion within highly curved fronts \citep[][M.~Freilich~2020, personal communication]{mackinnon2021nature}. Owing to its generic nature, the concept may also find relevance for other geophysical flows such as in the atmosphere or on other planets.

While presenting a framework for understanding sources and sinks of $Lq$ or ``potential momentum'' \citep[e.g.][]{Haynes:1987aa,Haynes:1990aa,marshallnurser1992pv} is beyond the scope of this study, one can nevertheless conceptually consider the theorem's application to the aforementioned flows by expressing Equation~\ref{EquationConservLq} for an axisymmetric vortex. This is done below for a vortex set at high latitudes and is followed by a brief discussion of the theorem's imprint on relative vorticity.

\subsection{Application to the axisymmetric vortex} \label{SectionAxiVortex}

We now consider the evolution of $Lq$ within an axisymmetric vortex in cyclogeostrophic and hydrostatic balance (\textit{i.e.}~GWB). We require frictional and diabatic effects to be weak enough such that this balance holds. This will consequently result in weak cross-frontal or radial fluid motion \citep{eliassen1951vortex}. Formally, we state that perturbations from the balanced state are small such that ${\mathbf{u} = \mathbf{\overline{u}} + \mathbf{u^{*}} \approx (0,\overline{v},0)}$ and ${b = \overline{b} + b^{*} \approx \overline{b}}$, where the overbar denotes mean quantities and asterisks ($^{*}$) denote deviations from this state. We neglect compressibility effects, make the Boussinesq approximation, and define $\lambda = -g\rho$, allowing the Ertel PV to be written as ${q = \mathbf{\omega_{a}} \cdot \nabla b}$, where $b = -g\rho/\rho_{o}$ is buoyancy and $\rho_{o}$ is a constant reference density. Finally, we set the meridional component of Coriolis to zero on the basis that the flow is sufficiently distant from the Equator.

We now express Equation~\ref{EquationConservLq} in cylindrical coordinates. The vertical component of absolute angular momentum is ${L = \mathbf{m_{a}} \cdot \mathbf{\hat{k}} = (\mathbf{m} + \mathbf{m_{\Omega}}) \cdot \mathbf{\hat{k}}}$, where ${\mathbf{m} = \mathbf{r_{c}} \times \mathbf{u}}$ and ${\mathbf{m_{\Omega}} = \mathbf{r_{c}} \times (\mathbf{\Omega} \times \mathbf{r_{c}}) }$ are the relative and planetary angular momentum. Together with ${\mathbf{r_{c}} = (r,\phi,z)}$, ${\mathbf{u} = (u,v,w)}$, and ${\mathbf{\Omega} = (0,0,f)}$, we find $ L = r\overline{v} + fr^2/2$. The Ertel PV for this vortex is ${q = (2\mathbf{\Omega} + \nabla \times \mathbf{u}) \cdot \nabla b}$. Together with GWB ${(f + 2\overline{v}/r) \partial_{z}\overline{v} = \partial_{r}\overline{b} = M^2}$,
we have ${q = (f + \overline{\zeta})N^2 - (f + 2\overline{v}/r) |\partial_{z} \overline{v}|^2}$, where we have neglected the horizontal vorticity owing to its smallness relative to other terms. Thus, the relative vorticity associated with the balanced state is ${\overline{\zeta} = (1/r)\partial_{r}(r\overline{v}) = \overline{v}/r + \partial_{r}\overline{v}}$ and vertical stratification is ${N^2 = \partial_{z}\overline{b}}$.
With these definitions in hand, Equation~\ref{EquationConservLq} becomes
\begin{eqnarray} \label{EquationConservGWB}
\frac{D\Pi}{Dt} = S_{\mathscr{F} + \mathscr{D} + \mathscr{P} + \mathscr{g} + \Omega},
\end{eqnarray}
where the potential momentum is
\begin{equation} \label{EquationDefPi}
\Pi = Lq = \left (r\overline{v} + \frac{fr^2}{2} \right )\left [ \left ( f + \overline{\zeta} \right )N^2 - \left (f + \frac{2\overline{v}}{r} \right ) |\partial_{z} \overline{v}|^2 \right ] = \frac{r^2\Phi}{2}
\end{equation}
and $S_{\mathscr{F} + \mathscr{D} + \mathscr{P} + \mathscr{g} + \Omega}$ represents sources and sinks of momentum due to frictional, diabatic, pressure, and gravitational sources, as well as Earth's rotation acting on the relative momentum vector (\textit{i.e.}~RHS of Equation~\ref{EquationConservLq}). $\Phi$ is the generalized Rayleigh discriminant and consists of barotropic and baroclinic components \citep{buckingham2020part1}:
\begin{equation} \label{EquationDefPhi}
\Phi = \underbrace{(f + 2\overline{v}/r ) ( f + \overline{\zeta} )N^2}_{barotropic}  - \underbrace{(f + 2\overline{v}/r )^2 |\partial_{z}{\overline{v}}|^2}_{baroclinic} = \chi^2N^2 - M^4
\end{equation}
where ${\chi^2 = (f + 2\overline{v}/r ) ( f + \overline{\zeta} )}$ is the generalized Rayleigh discriminant for barotropic vortices \citep{Kloosterziel:1991aa,mutabazi1992vortex}.
As in the introduction, we define gradient Rossby, gradient Richardson, and curvature numbers as ${\mathrm{Ro} =  {\overline{\zeta}}/{f}}$, ${\mathrm{Ri} = {N^2}/{|\partial_{z}{\overline{v}}|^2}}$, and ${\mathrm{Cu} = {2\overline{v}}/{fr}}$, allowing us to also write the potential momentum as
\begin{equation} \label{EquationDefPiNonDim}
\Pi = Lq = \left ( \frac{f^2 N^2 r^2}{2} \right ) \Phi',
\end{equation}
where 
\begin{equation} \label{EquationDefPhiNonDim}
\Phi' = L'q' = \left (1 + \mathrm{Cu} \right ) \left ( 1 + \mathrm{Ro} \right ) - \left (1 + \mathrm{Cu} \right )^2 \cdot \mathrm{Ri}^{-1}.
\end{equation}
is a nondimensional form of the generalized Rayleigh discriminant, $L' = 1 + \mathrm{Cu}$ is a nondimensional form of absolute angular momentum, and $q'$ is the nondimensional Ertel PV. Expanding Equation~\ref{EquationConservGWB}, we find
\begin{equation} \label{EquationConservPhi}
{\frac{D\Phi}{Dt} = 2{S_{\mathscr{F} + \mathscr{D} + \mathscr{P} + \mathscr{g} + \Omega}} - \frac{2u}{r}\Phi }.
\end{equation}
Thus, in the absence of sources and sinks of potential momentum (${S_{\mathscr{F} + \mathscr{D} + \mathscr{P} + \mathscr{g} + \Omega} = 0}$) and assuming no cross-frontal motion (${u = 0}$), the stability of the vortex is constant: $D\Phi/Dt = 0$. However, if a fluid parcel moves radially ($u \ne 0$)--even in the absence of sources and sinks of potential momentum--there must be a corresponding change in the stability of the flow.
This point above contrasts with statements made by \citet{buckingham2020part1}. In particular, in their discussion, \citet{buckingham2020part1} speculated that $\Phi$ and $\Phi'$ might be conserved quantities following fluid parcels within meandering fronts and vortices. As evidenced by Equation~\ref{EquationConservPhi}, this statement is not entirely true. We can expect cross-frontal motion as a fluid parcel is advected along its path in a meandering flow \citep{bower1989pv,samelson1992meandering} such that $\Phi$ is not conserved. Additionally, $D\Phi/Dt = 0$ does not imply $D\Phi'/Dt = 0$, except if $N^2$ and $r^2$ do not change following a fluid parcel (cf.~Equation~\ref{EquationDefPiNonDim}). These conditions can be met within the axisymmetric vortex ($\partial_{\phi} \overline{b} \approx 0$ and $u \approx 0$) but will not generally be met following fluid parcels within a meandering front. \textbf{These points therefore correct and/or qualify statements made by \citet{buckingham2020part1,buckingham2020part2}.} It also demonstrates that, to the extent that $\Phi$ can be used as a predictor for along-isopycnal motion, the theorem may find use in understanding parcel motion in curved baroclinic fronts.

In general, application of the vorticity theorem requires detailed knowledge of each of the terms represented by ${S_{\mathscr{F} + \mathscr{D} + \mathscr{P} + \mathscr{g} + \Omega}}$. For example, for a polar vortex located under ice but away from seamounts, $\mathbf{\Omega_{o}} \times \mathbf{m} \approx 0$, the gravitational torque is zero, frictional and diabatic terms are likely nonzero, and depending upon the topography of the ice, gradients in pressure and the corresponding pressure torque could also be nonzero (K.~Nicholls, 2020; personal communication). In such a situation, specification of these source/sink terms can be non-trivial. However, a unique situation exists in which all of these terms are approximately zero. This case is considered below.

\subsubsection*{Submesoscale and polar mesoscale vortices} \label{SectionVortexExample}

\begin{figure}[t]
\begin{centering}
\noindent\includegraphics[width=18pc,angle=0]{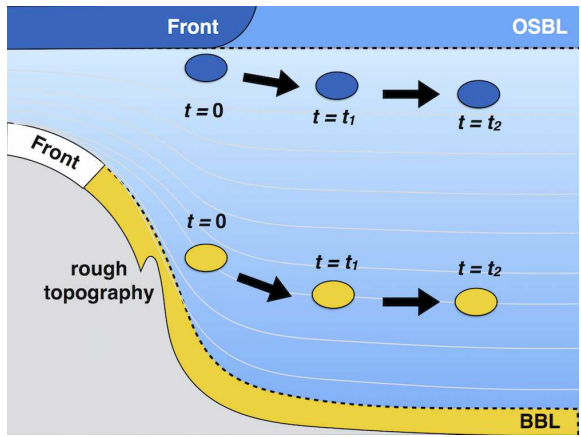}
  \caption{Concept of the generation and evolution of small-scale, coherent vortices in the ocean interior, adapted from \citet{buckingham2020part2}. Although other formation mechanisms are possible, here vortices are depicted as forming from a front in the ocean surface boundary layer~(OSBL) and flow-topography interaction in the bottom boundary layer~(BBL), with each color denoting a different boundary layer fluid (blue=OSBL, yellow=BBL). In general, in regions of low stratification, conservation of $Lq$ tends to generate flow with anticyclonic vorticity. However, instances exist when cyclones are produced, such as by abrupt topography. Immediately following formation, the cyclone has large relative vorticity (at time $t = 0$). However, due to reduced core stratification and corresponding low Richardson numbers, the cyclone must modify shears, stratification, and centripetal accelerations to ensure ${Lq > 0}$. We propose that this occurs via energy loss resulting from symmetric instability. Eventually, the cyclone becomes stable ($t = t_{2}$) but its final vorticity is reduced. In contrast, the anticyclone, being weakly or marginally stable, is not as significantly affected and maintains its energy for a greater number of inertial periods (${t = 0,t = t_{1},}$~...~${,t = t_{2}}$). A time scale for the decay of the cyclone is given in the main text and is inversely proportional to the square root of $-\Phi'$~(cf.~Equation~\ref{EquationDefPhiNonDim}).}\label{f5}
\end{centering}
\end{figure}

Coherent baroclinic vortices with radii of $1$-$10$~km are believed to be ubiquitous in the world's oceans and are generated in close proximity to boundaries (Figure~\ref{f5}). Despite that they include polar \textit{mesoscale}  vortices, this class of vortices has been termed ``{submesoscale coherent vortices}'' (SCVs) owing to their comparable size, intensity and stratification as those found at mid-latitudes \citep{mcwilliams1985scv}. In particular, they are marked by significantly reduced vertical stratification and unique watermass properties. They appear to have been first documented as a distinct water-mass in hydrographic observations near the Bahamas in the Atlantic \citep{mcdowell1978meddy}. Since, however, they are now regarded as ubiquitous in the global oceans \citep[\textit{e.g.}][]{mccoy2020scv}.

The prevailing understanding is that SCV formation can happen via a horizontal shear instability, for example, past topography or within an island wake \citep{Barkley1972,Stegner2014bookchapter,gula2016nature,MacKinnon:2019aa,srinivasan2019topo}, vertical shear or mixed layer baroclinic instability \citep{haine1998gravitational,eldevik2002baroclinic,boccaletti2007mixed}, as well as convection of boundary layer fluid \citep{legg2001labrador,bosse2016scv}. This set of boundary layers include the ocean surface boundary layer (OSBL), bottom boundary layer (BBL), and ice-ocean boundary layer (IOBL). At formation, vortices will typically spin up in such a way that an anticyclone is preferred. This occurs principally due to conservation of $q$ (with $fq > 0$ initially) \citep{spall1995front,thomas2008intrathermocline} but conservation of $L$ nevertheless plays an important role owing to centripetal accelerations experienced by fluid parcels. It follows from the preceding derivation that Equation~\ref{EquationConservLq} applies, so long as the fluid is continuous in the azimuth direction such that $L$ can be appropriately defined.

One of the consequences of Equation~\ref{EquationConservLq} is that geophysical vortices which reside away from boundaries and perturbations in Earth's geopotential approximately conserve the product of $L$ and $q$. In the absence of sources and sinks of potential momentum (${S_{\mathscr{F} + \mathscr{D} + \mathscr{P} + \mathscr{g} + \Omega} = 0}$), we can safely approximate cross-frontal motion within a vortex as zero ($u \approx 0$). By virtue of Equation~\ref{EquationConservPhi}, this also implies that the Rayleigh discriminant $\Phi$ is conserved: ${D\Phi/Dt \approx 0}$. To understand the consequences of these statements, we revisit the generation and evolution of SCVs.

While the details of this evolution remain unclear without corroborating laboratory or model support, the following arguments are reasonable. Now, the low-stratified boundary layer fluid will be characterized by $\mathrm{Ri} \sim 1$. If this fluid is trapped within the vortex core,\footnote{One measure of this trapping is the ratio of the azimuthal speed $|\overline{v}|$ to the wave propagation speed $c$ \citep{samelson1992meandering,chelton2011eddies}. Given an internal wave speed $c = NH$ and maximum velocity scale $v_{m}$, this condition is equivalent to ${v_{m}/(NH) > 1}$.} then the fluid with low $\mathrm{Ri}$ will persist even as the vortex subducts or is advected away from the boundary (Figure~\ref{f5}). It immediately follows that the results from \citet{buckingham2020part2} apply. That is, symmetric instability will be active within cyclonic vortices, while anticyclonic vortices will remain marginally stable, decaying over long time scales due to weak inertial-symmetric instabilities. A timescale for the decay of the cyclonic vortex can be estimated from the growth rate of symmetric disturbances: ${T = 2\pi/\sigma}$, where ${\sigma = f(-\Phi')^{1/2}}$ is an approximate growth rate of symmetric disturbances under a simplified axisymmetric vortex model \citep[][Appendix~A]{buckingham2020part1}. An analytical time scale for the anticyclone is unknown to the author, although anecdotally it has been estimated at months to years based on watermass properties \citep{mcdowell1978meddy,mcwilliams1985scv}.\footnote{\citet{mahdinia2017jfm} observe a timescale greater than $50$~eddy ``turnaround times'' within stability analysis of Gaussian vortices.}

One point not considered above is the advection of the vortex into a different ocean environment. In this case, the vortex must alter barotropic and baroclinic components of $\Phi$ so as to keep $\Phi$ constant. Thus, vertical stratification, vorticity, vertical shear, and curvature (cf.~Equation~\ref{EquationDefPhi}) must change, where vorticity is related to curvature through ${\mathrm{Ro} = \partial_{r}\overline{v}/f + \mathrm{Cu}/2}$. If, however, the vortex is unable to alter these parameters quickly enough, energy will be dissipated until an equilibrium is reached. In non-dimensional form (cf.~Equations~\ref{EquationDefPiNonDim}~and~\ref{EquationDefPhiNonDim}), we find that $\mathrm{Ro}$, $\mathrm{Ri}$, and $\mathrm{Cu}$ must change in some way so as to conserve the non-dimensional form of $Lq$ (\textit{i.e.}~$\Phi'$). This can be seen as a form of cyclo-geostrophic adjustment \citep{stegner2004part1}. Note: Equation~\ref{EquationDefPiNonDim} applies but ${D/Dt(N^2) \ne 0}$.

\subsection{Comment on the distribution of vorticity in the oceans}

We close this study with one final comment. It is generally understood that the distribution of vorticity as measured at small horizontal scales in the oceans has two characteristics. Expressed in non-dimensional form, the distribution of vorticity $\zeta/f$ is positively skewed (mostly cyclonic) in straight fronts and negatively skewed (mostly anticyclonic) in eddying or vortex flows. This has been noted, for example, in upper ocean observations and model simulations when examined at submesoscale resolution \citep{rudnick2001vorticity,shcherbina2013statistics,buckingham2016seasonality}. For reference, we refer the reader to \citet[][Figure~5b]{shcherbina2013statistics}. While the former can be rationalized in terms of PV conservation \citep{HoskinsBretherton1972}, the latter has not fully been explained. Here, we offer a simple explanation for this observation.

Note that requiring positive potential momentum $Lq > 0$ for all time is equivalent to requiring balanced flow ($\Phi > 0$) for all time. In this case, \textit{$D/Dt(Lq) = 0$ together with an initial positive state $Lq > 0$ places constraints on the sign of $\Phi$ and determines the distribution of relative vorticity in the oceans.} This is analogous to how $Dq/Dt = 0$ together with an initial positive state ${fq > 0}$ determines the distribution of relative vorticity at straight fronts \citep{buckingham2016seasonality}. Another way to state this is that the statistics of relative vorticity are determined by the possible set of Rossby numbers which ensure the stability discriminant is positive: $\Phi > 0$ or $\Phi' > 0$. If one requires that Equation~\ref{EquationDefPhi}~or~\ref{EquationDefPhiNonDim} be positive and solves for the set of Rossby numbers which ensure this is true, the negative skewness discussed above will emerge at low $\mathrm{Ri}$ \citep[e.g.~Figure 13 of][]{buckingham2020part2}. While this skewness is revealed due to the effect of curvature on the Ertel PV $q$, helping to stabilize anticyclonic flow while de-stabilizing cyclonic flow through a tilting of the vorticity vector, $L$ plays an important role in bounding anticyclonic vorticity.

\section{Limitations} \label{SectionLimitations}

In this study, we have stressed the vertical component of absolute angular momentum. As a result, Equation~\ref{EquationConservLq} cannot be used predict $Lq$ since $L$ is coupled to the other components of $\mathbf{m_{a}}$ through ${\mathbf{\Omega} \times \mathbf{m}}$. As stated above, this term arises from Earth's rotation and tilts the absolute angular momentum vector away from the vertical, reducing the magnitude of $L$. This torque is greatest at the Equator and zero at the poles. Such a case might arise when a tropical vortex wobbles about its vertical axis of rotation--\textit{i.e.}~when the relative angular momentum $\mathbf{m}$ is not entirely vertical. This does not make Equation~\ref{EquationConservLq} incorrect but simply limits its utility in certain cases.

The author has not attempted to reframe the theorem for the $\beta$-plane, although this would indeed be a useful exercise since intense vortices and tropical instability waves are found in the vicinity of the Equator \citep{marchesiello2011tiw,holmes2013pvtiw,iury2021ant}. Progress in this area might be made by examining the work of \citet{Grimshaw1975beta} and \citet{Kloosterziel:2017aa}. In conclusion, these restrictions--\textit{i.e.}~the (1)~restriction to the $f$~plane and (2)~growing importance of ${\mathbf{\Omega} \times \mathbf{m}}$--are the two main limitations to the theorem. A third limitation might also be that the fluid must be continuous in the azimuth direction in order to properly define $L$, but at small horizontal scales this is easily achieved.

It may be worth noting that the $f$~plane approximation together with angular momentum conservation principles have previously been successful for investigating tropical cyclone dynamics \citep{houze1993text}, thus indicating that these limitations--which grow for vortices closer to the Equator--may not be so severe. It is probable that a more elegant intersection of these three principles (\textit{i.e.}~Figure~\ref{f1} but where $L$ is replaced by $\mathbf{m_{a}}$), will be presented in the future.

\section{Conclusions} \label{SectionConclusions}

In this study, we have presented a conservation equation valid on the $f$~plane. It is an extension of Ertel's PV theorem to flow at small horizontal scales such that absolute angular momentum is appropriately conserved. The combination of absolute angular momentum conservation together with Ertel's theorem has implications for the motion of fluid parcels. In particular, we discovered two important consequences of the theorem: (1)~shear, stratification, and centripetal accelerations are modified in concert in an effort to conserve $Lq$ and (2)~asymmetry or skewness in the distribution of relative vorticity results directly from this conservation principle, permitting the occurrence of stable anticyclonic curved flow, while limiting the occurrences of cyclonic curved flow at low Richardson numbers \citep{buckingham2020part1,buckingham2020part2}. That is, if $Lq > 0$ initially, then $D/Dt(Lq) = 0$ has important consequences for the range of vorticity values seen at small horizontal scales in the ocean. While this may find clearest application in explaining why SCVs are overwhelmingly anticyclonic \citep{mcdowell1978meddy,riser1986scv,DAsaro:1988aa,Bane:1989aa,timmermans2008arctic,zhao2014arctic}, the theorem will also find use in understanding Lagrangian parcel motion within highly curved baroclinic fronts.

The topic of absolute angular momentum conservation has received little attention in oceanography texts, while this same topic has received considerable attention in the atmospheric literature \citep{holton1992text,piexoto_oort1992physics,barnes1983momentum,bell1994momentum}. This appears to be due to the presence of continental boundaries in the ocean but which are absent in the atmosphere \citep{griffies2004text}, causing PV rather than absolute angular momentum to be a more universally conserved quantity at large horizontal scales \citep{pedlosky1987gfd}. An exception may be in the Southern Ocean, where obstacles to zonal flow are absent \citep{straub1993ang}. However, for small-scale geophysical flows in which centripetal accelerations are present and Earth's rotation plays a dynamically important role,\footnote{A helpful point here is the following: Earth's rotation imparts angular momentum to fluid parcels and can limit centripetal accelerations present in anticyclonic flows through the constraint, $L' = 1 + \mathrm{Cu} > 0$, assuming $\mathrm{Cu} > -1$ initially. This is analogous to how Earth's rotation imparts vorticity to fluid parcels and limits horizontal shear present in anticyclonic flows through the constraint, $1 + \mathrm{Ro} > 0$.} the conservation of absolute angular momentum finds its place. Submesoscale and polar mesoscale flows are ideal examples in which such a conservation principle may apply. It is in the context of these phenomena that the combined theorem should find greater use.

\acknowledgments

This study was made possible by an Individual Fellowship from the European Commission on centrifugal instability (2019-2020). The derivation was made in isolation during a two-month period (December 2020-January 2021) following the fellowship. Although numerous individuals might be mentioned, the author wishes to acknowledge the following individuals: George Nurser, Guillaume Roullet, Stephen Griffies, Keith Nicholls, Greace Crystle, Jonathan Gula, Xavier Carton, Charly DeMarez, and Thomas Meunier. Of these, George Nurser made an important comment to the author regarding PV conservation (May 2015); Charly DeMarez served as an important sounding block as these ideas were forming (January-February 2020) and should be acknowledged; and Greace Crystle patiently and kindly listened to the author's musings and frustrations, playing a quiet but significant role in this study. Comments by several anonymous reviewers are also acknowledged. Subsequent conversations with other scientists (\textit{e.g.}~Amit Tandon, Glenn Flierl, and Bruno Deremble) have been very helpful. This article is dedicated to Alex~Stegner, who mentored numerous individuals in this field, and we would like to highlight the recent work of Suraj Singh and Manikandan Mathur which is closely related to this problem in the viscous context.

%
%


 \bibliographystyle{ametsoc2014}
 \bibliography{references}



\end{document}